\documentclass{ecai}
\usepackage{times}
\usepackage{graphicx}
\usepackage{latexsym}
\usepackage{balance}
\usepackage{amsmath}
\usepackage{newtxtext}       %
\usepackage{newtxmath}       
\usepackage{caption}
\usepackage{subcaption}

\makeatletter
\def\blfootnote{\xdef\@thefnmark{}\@footnotetext}
\makeatother
\begin{document}

\title{Detecting Signatures of Early-stage Dementia with Behavioural Models Derived from Sensor Data}


\author{
    Rafael Poyiadzi\textsuperscript{*}\institute{University of Bristol, England, email: rp13102@bristol.ac.uk} 
    \and
    Weisong Yang\textsuperscript{*}\institute{University of Bristol, England, email: ws.yang@bristol.ac.uk} 
    \and
    Yoav Ben-Shlomo \institute{University of Bristol, England, email: y.ben-shlomo@bristol.ac.uk}
    \and
    Ian Craddock \institute{University of Bristol, England, email: ian.craddock@bristol.ac.uk} \\
    Liz Coulthard \institute{University of Bristol, England, email: elizabeth.coulthard@bristol.ac.uk} 
    \and
    Raul Santos-Rodriguez \institute{University of Bristol, England, email: enrsr@bristol.ac.uk} 
    \and
    James Selwood\institute{University of Bristol, England, email: james.selwood@bristol.ac.uk} 
    \and 
    Niall Twomey \institute{Cookpad Ltd, and University of Bristol, England (honorary), email: niall-twomey@cookpad.com}
}

\maketitle
\bibliographystyle{ecai}

\abstract{There is a pressing need to automatically understand the state and progression of chronic neurological diseases such as dementia. The emergence of state-of-the-art sensing platforms offers unprecedented opportunities for indirect and automatic evaluation of disease state through the lens of behavioural monitoring. This paper specifically seeks to characterise behavioural signatures of mild cognitive impairment (MCI) and Alzheimer's disease (AD) in the \textit{early} stages of the disease. We introduce bespoke behavioural models and analyses of key symptoms and deploy these on a novel dataset of longitudinal sensor data from persons with MCI and AD. We present preliminary findings that show the relationship between levels of sleep quality and  wandering can be subtly different between patients in the early stages of dementia and healthy cohabiting controls.}

\blfootnote{\textsuperscript{*} Authors contributed equally}


\section{Introduction} \label{introduction}

Dementia is a progressive neurological condition affecting cognition and behaviour with a significant impact on activities of daily living. It is one of the major causes of disability and dependency among elderly population with approximately 50 million people with the disease world wide. Alzheimer's disease (AD) is the most common cause of dementia. Mild cognitive impairment (MCI) refers to a decline that has minimal impact on activities of daily living \cite{grundman2004mild}. Patients with MCI can continue to live independently \cite{mckhann2011diagnosis}. Not all patients with MCI will convert to dementia, but they are at increased risk. MCI has an annual conversion rate to AD of between 5 and 10\% \cite{mitchell2009rate}.

Making an accurate, reliable diagnosis of dementia is a challenge. Patients may find cognitive tests stressful, which impacts their performance. This may be exacerbated in a new, unknown, clinical environment with an unfamiliar clinician. Performance on cognitive tests also does not give an indication of how someone is managing in the real world. Although diagnosis may include bio-markers measurement, in addition to this being invasive and expensive, these do not have adequate temporal resolution to understand disease progression and deterioration since AD has high variance across patients.

Some symptoms are prevalent across dementias. 
Confusion can lead to unpredictable behaviour and activity abandonment during normal daily routines. Interestingly, behavioural interactions between persons with dementia (PwD) and their carers/family are also affected by disease progression and efforts by PwD to allay confusion and uncertainty often result in them persistently seeking the company of persons whom they trust. 
In this work we intend to address \textit{shadowing} \cite{lilly2011illumination} (persistently staying in the company of a trusted carer/family), \textit{wandering} \cite{lyketsos2011neuropsychiatric} (going from room to room with nonspecific intent) and \textit{disturbed sleep} \cite{hita2012disturbed} (a common symptom for PwD).

Progression of dementia and MCI are characterised by an increase in symptoms that will affect normal daily activity and behaviour. It is important to remember that these symptoms affect normal daily activity and behaviour. In understanding the dependence between normal behaviour and the expression of symptoms, an opportunity to understand and quantify disease state indirectly via behavioural change is exposed. 
Modern sensing technology offers promise in monitoring daily behaviour, extracting symptom expression rates and measurement of state and progression of the disease \cite{de2017feasibility,stack2018identifying}. In contrast to normal clinical evaluation that occur in foreign environments, ours is achieved from the patients' residences. 


This paper outlines novel computational behavioural analysis algorithms for modelling disease state and progression. The potential for detecting unseen behavioural bio-markers of the early stages of dementia is exposed by analysis on the longitudinal, in-home sensor data. 
Section \ref{related_work} reviews prior work on smart homes and dementia. Our data collection and modelling procedures are outlined in Section \ref{methods}. Results are presented in Section \ref{experiments} and we conclude in Section \ref{conclusion}.
\section{Related Work} \label{related_work}
Numerous smart systems have been designed and developed to monitor the well-being and health status of elders, such as the GATOR Tech Smart House \cite{helal2005gator}, the AWARE home \cite{abowd2000living}, the Microsoft’s EasyLiving project \cite{brumitt2000easyliving} and the MavHome Project \cite{cook2006health}. In the dementia domain, \cite{li2017supporting} introduced a support system to aid doctors diagnosing dementia. Participants only needed to perform a selection of Instrumental Activities of Daily Living (IADL) in a smart home environment. However, the main limitation is that this approach relies on data collected in a laboratory environment and not in a real world setting. 

Along these lines, \cite{akl2015autonomous} undertook a study on data acquired by The ORegon Centre for Aging and TECHnology (ORCATECH) at the Oregon Health and Science University, that investigated the walking speed and general activity in the homes of participants in order to detect dementia. The extension presented in \cite{akl2015unobtrusive} modelled the presence of the participants in different rooms using Poisson processes to demonstrate statistical differences between different states of cognition. 
Localisation has been found to be a common ground for different works. For instance \cite{kearns2008ultra} and \cite{schwarz2005accuracy} use an Ultra Wideband device to try to measure wandering behaviours. However, Radio Frequency (RF) identification may be limited by its detection range since unpowered RF devices operate only over a meter's distance. Also, \cite{d2012indoor} uses a relatively complex sensor network consisted of RF tracking and motion and heading sensors for monitoring and studying behavioural patterns of patients with dementia. 

We build upon the existing literature by focusing on data collected in the wild that, while more challenging to analyse, also provides a richer and pervasive perspective on the subjects’ behaviour.
Additionally, our aim is to automatically detect behavioural symptoms that may suggest \textit{early} dementia. Finally, we want go beyond the analysis of a single participant and incorporate the monitoring of interactions within the home environment as a key element to understand the progression of the disease.

\section{Methods} \label{methods}
This section introduces our data collection and modelling pipelines. 

\subsection{Data collection}

Ethical approval was secured from the Carmarthen Research Ethics Committee and the Health Research Authority (HRA) to record sensor data from the homes of people who have received a diagnosis of dementia or have MCI for up to 12 months using the SPHERE \cite{zhu2015bridging,woznowski2017sphere,diethe2018releasing} system. Informed consent was obtained to analyse this data to determine links between behavioural patterns, residents of the home and disease state.  Participants are recruited from North Bristol NHS Trust, the Dementia Wellbeing Service in Bristol and the Research Institute for the Care of Older People (RICE) Centre in Bath. The SPHERE sensor system is installed into their homes \cite{zhu2015bridging,woznowski2017sphere}. 


The key sensors that we use in the analysis of this work are accelerometers \cite{elsts2018guide,fafoutis2017designing,vafeas2020wearable}. These are set record at a rate of 20 Hz. This data is broadcast to gateways by means of radio links. Six gateways are installed in every house. These simultaneously record accelerometer readings and the Received Signal Strength Intensity (RSSI). 


In-home localisation is a key task in this work that leverages the RSSI data \cite{michal2019, michal2018, michal2017}. Labelled data are acquired from a `walkthrough' script that is performed by the technician upon installation. Each room is visited while holding the wearables and RSSI and wearable data are simultaneously recorded. The real-time location of the technician is recorded with an annotation app. 

To regress from sensor data to localisation and activity predictions, we extract basic features from windowed data ($5$ seconds length with overlap of $2.50$ seconds following \cite{twomey2018comprehensive}). Features include: \emph{mean}, \emph{standard deviation} (std), \emph{max}, \emph{min}, \emph{diff} (first-order difference operator) and the count of missing values. 

We can complement labelled data by introducing domain knowledge into the analysis by suggesting reasonable locations based on calculable context. For example, if predicting `sleep' at 3AM it is likely that the resident is in the bedroom.



\subsection{Modelling considerations}


\subsubsection{Covariate Shift}
We account for our expectations of covariate shift (due to disease progression) with Maximum Mean Discrepancy (MMD) \cite{huang2007correcting}. 
%
%
In mathematical terms we assume that training and testing distributions differ, \textit{i.e.} $P_{tr}(x) \neq P_{te}(x)$, but that conditional distributions remain the same $P_{tr}(y|x) = P_{te}(y|x)$. MMD offers an elegant approach to quantify and match distributional shift and we follow the methodology outlined by \cite{huang2007correcting} in our work. 

\subsubsection{Semi-Supervised Learning}
Since only a small portion of labelled data are available, we leverage supervised ($\mathcal{L}_{SL}$) semi-supervised ($\mathcal{L}_{SSL}$) objectives to make efficient use of data. 
The supervised objective ($\mathcal{L}_{SL}$) is the traditional conditional log-likelihood that one optimises for in Conditional Random Fields (CRF) \cite{sutton2012introduction}. The data and labels for this are obtained from the walkthrough that was introduced in the previous section. 
%
%
CRF model sequence probabilities as:
\begin{equation}
    \label{eq:crf}
    P_{\theta'}(\boldsymbol{y} | \boldsymbol{X}) = Z(\boldsymbol{X})^{-1} \cdot \exp \left(\sum_{t=1}^{T} \phi_{t}\left(x_{t}, y_{t}, y_{t-1}\right)\right)
\end{equation}
%
%
where $\phi_t$ are the log-potentials and $Z(\boldsymbol{X})$ is the partition function. In our case the log-potentials decompose, to \emph{emission} function and \emph{transition} function, as follows: $\phi_{t}\left(x_{t}, y_{t}, y_{t-1}\right)$ = $u\left(x_t, y_t\right) \cdot \tilde{\tau}(y_{t}, y_{t-1})$, where: 

\begin{equation}
    \label{eq:switching}
    \tilde{\tau}(y_t, y_{t-1}) =
    \left\{ \begin{array}{ll}
        \tau(y_t, y_{t-1}), & \alpha(t)>=\epsilon'  \\
        \mathbb{I}_{y_t = y_{t-1}}, &\text{otherwise}
    \end{array} \right.
\end{equation}
where $\alpha(t)$ is the activity level of the participant at time $t$ (calculated as the average absolute jerk of the tri-axial signals) and $\epsilon'$ denotes a threshold, below which we assume there is no movement. This forbids location transitions unless a certain activity potential is surpassed.  

The emission potentials are modelled with a 4-layer fully-connected neural network, $u(x,y) = h(x)_y$, \textit{i.e.} the $y$-th element of $h(x)$. The semi-supervised objective ($\mathcal{L}_{SSL}$) constructed by iteratively using the most likely label sequence as temporary targets, \textit{i.e.} $\boldsymbol{y}^* = \arg \max_{\boldsymbol{y} \in \mathcal{Y}} P(\boldsymbol{y} | \boldsymbol{X})$ \cite{alex2020torchstruct}.

\section{Findings} \label{experiments}
In this section we present findings relevant to \emph{sleeping disturbance}, \emph{wandering} and \emph{shadowing}. We illustrate the analysis using $2$ different houses that are inhabited by either a PwD, or by someone who has been founds to have MCI, and one partner. Data collection is ongoing, so these are preliminary findings and some houses have significantly more data than others. The analyses below result from localisation and activity level predictions that were trained and validated on the walkthrough data. 

\subsection{Shadowing}
\begin{figure*}
\centering
\begin{subfigure}{0.3\linewidth}
  \centering
  \includegraphics[width=1.\linewidth]{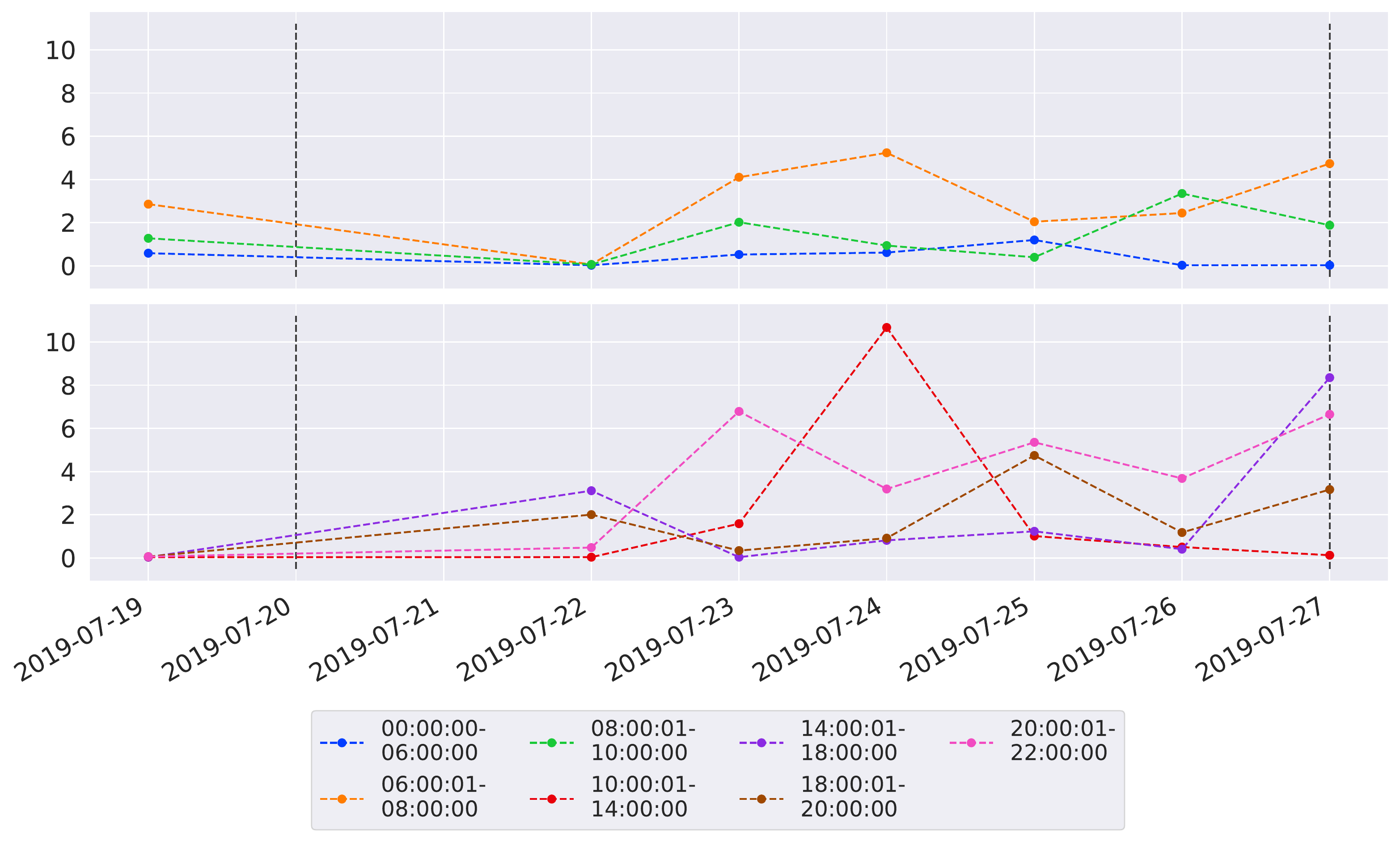}
  \caption{House A}
  \label{fig:sub1}
\end{subfigure}~
\begin{subfigure}{0.3\linewidth}
  \centering
  \includegraphics[width=1.\linewidth]{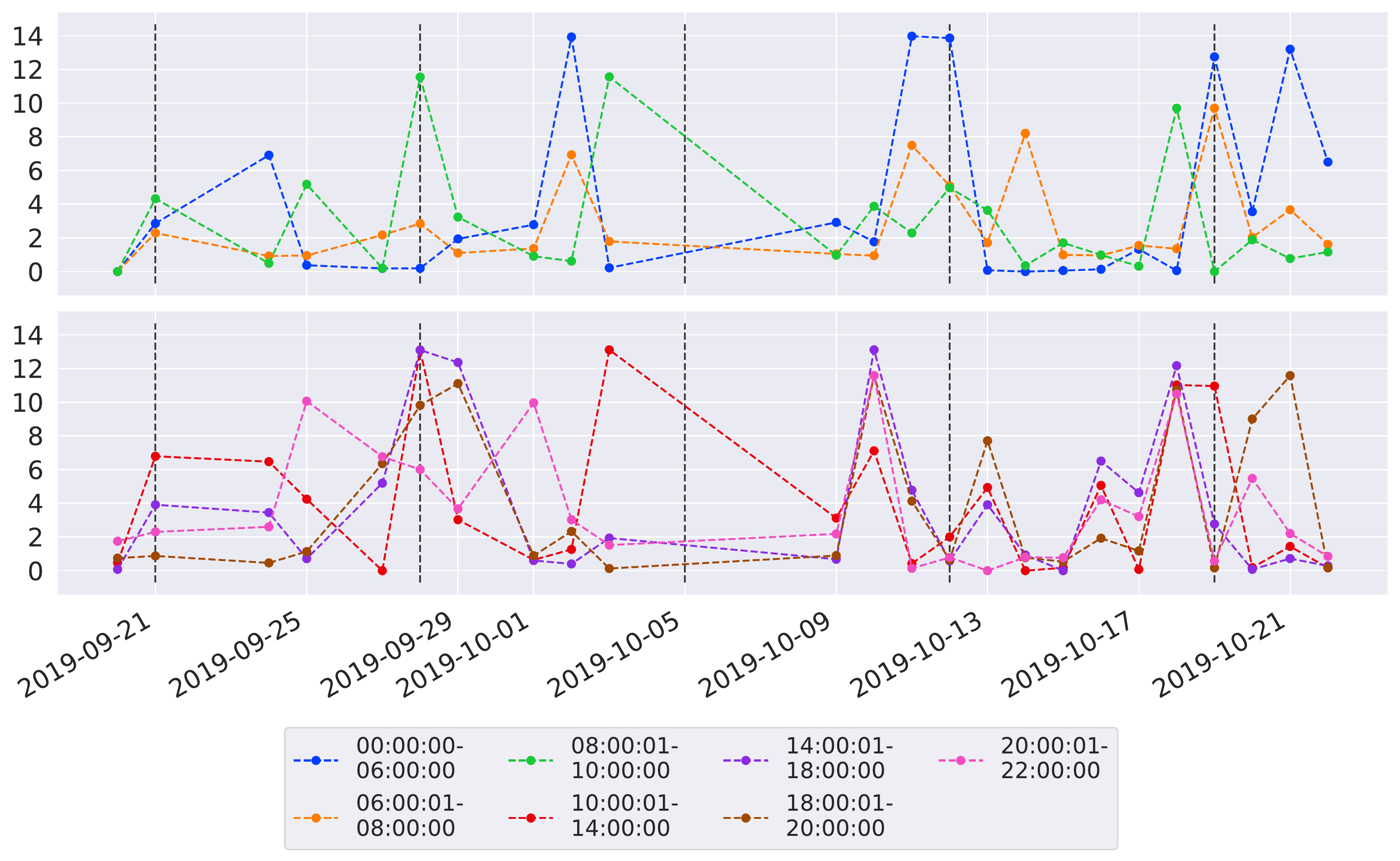}
  \caption{House B}
  \label{fig:sub2}
\end{subfigure}
\caption{MI derived for wandering, separated in different segments of the day. 
}
\label{fig:mi}

\centering
\begin{subfigure}{.6\linewidth}
  \centering
  \includegraphics[width=1.\linewidth]{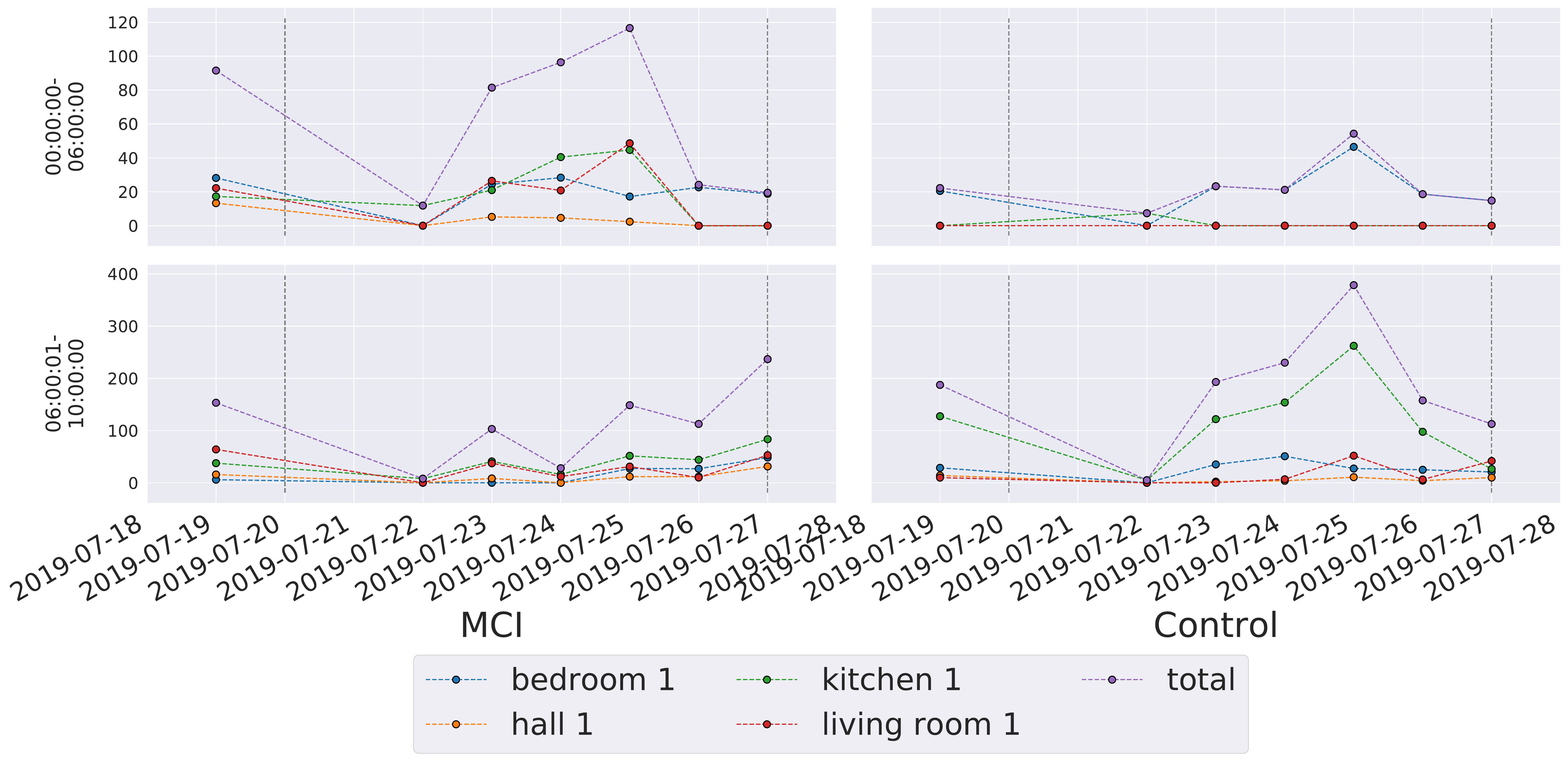}
  \caption{House A}
  \label{fig:acc3011}
\end{subfigure} \\
\begin{subfigure}{.6\linewidth}
  \centering
  \includegraphics[width=1.\linewidth]{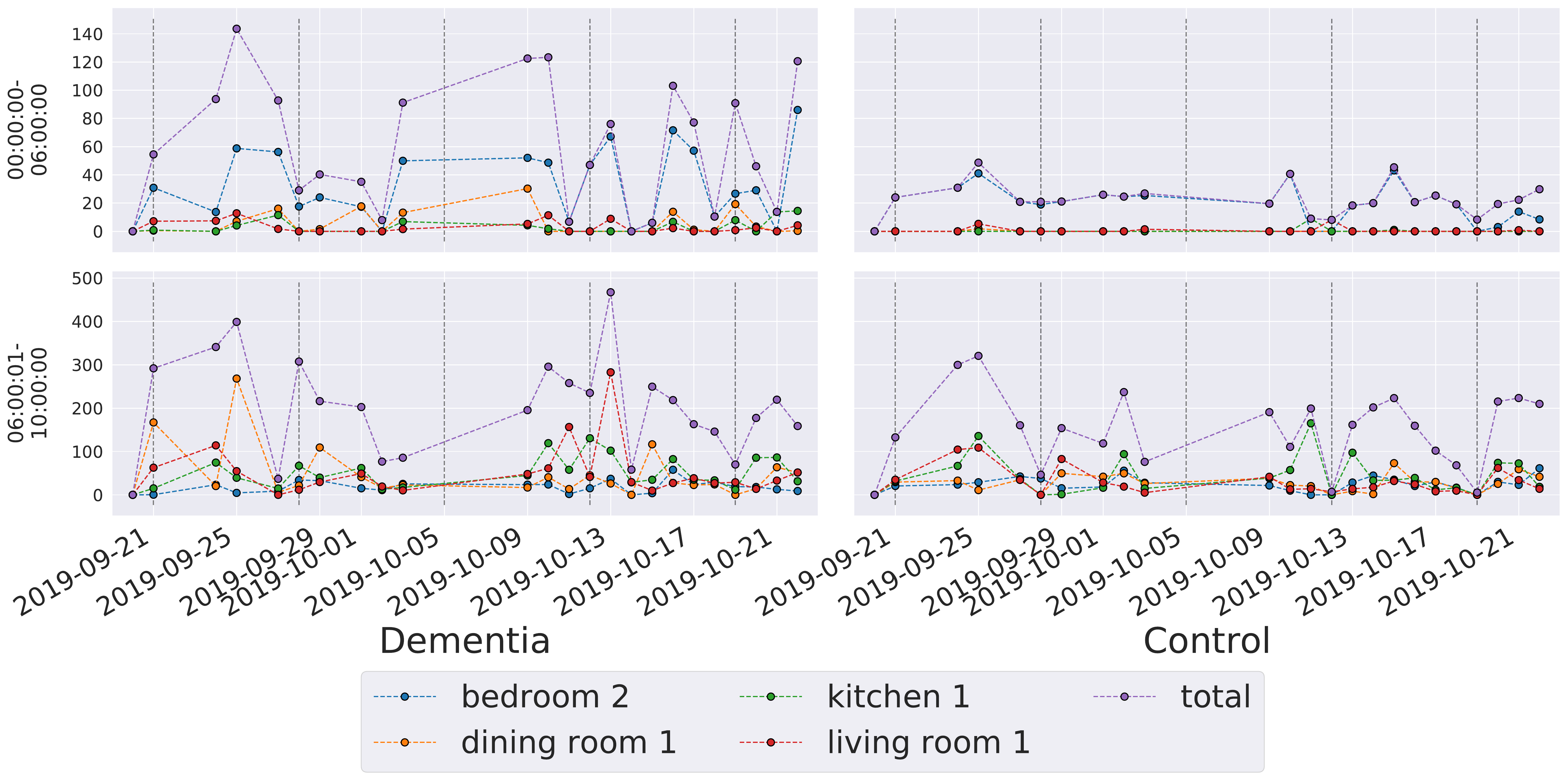}
  \caption{House B}
  \label{fig:acc5110}
\end{subfigure}
\caption{The total activity per room for House A (upper) and House B (lower).}
\label{fig:activities}

%
%
\centering
\begin{subfigure}{.3\linewidth}
  \centering
  \includegraphics[width=\linewidth]{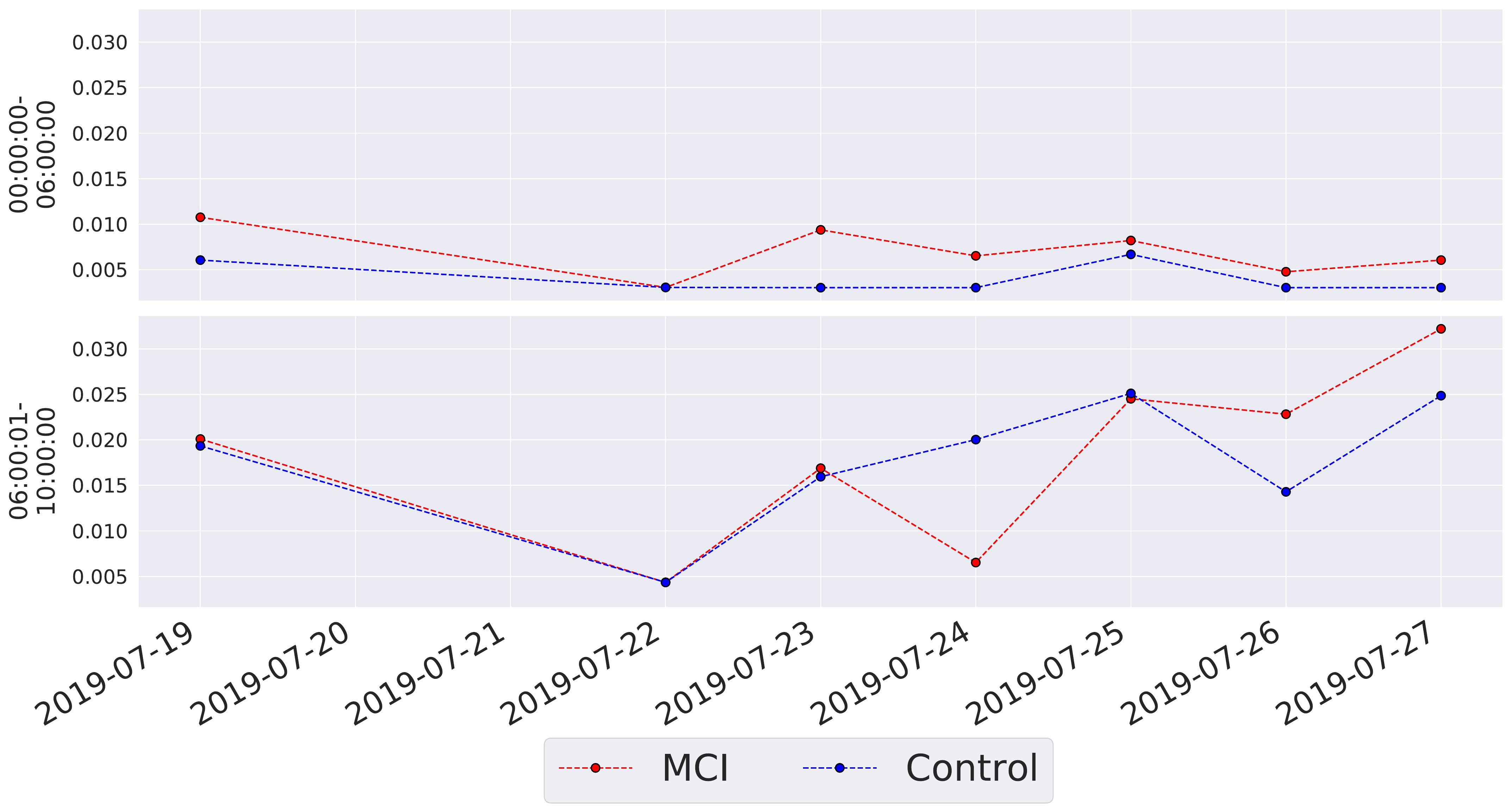}
  \caption{House A}
  \label{fig:entropy3011}
\end{subfigure}~
\begin{subfigure}{.3\linewidth}
  \centering
  \includegraphics[width=\linewidth]{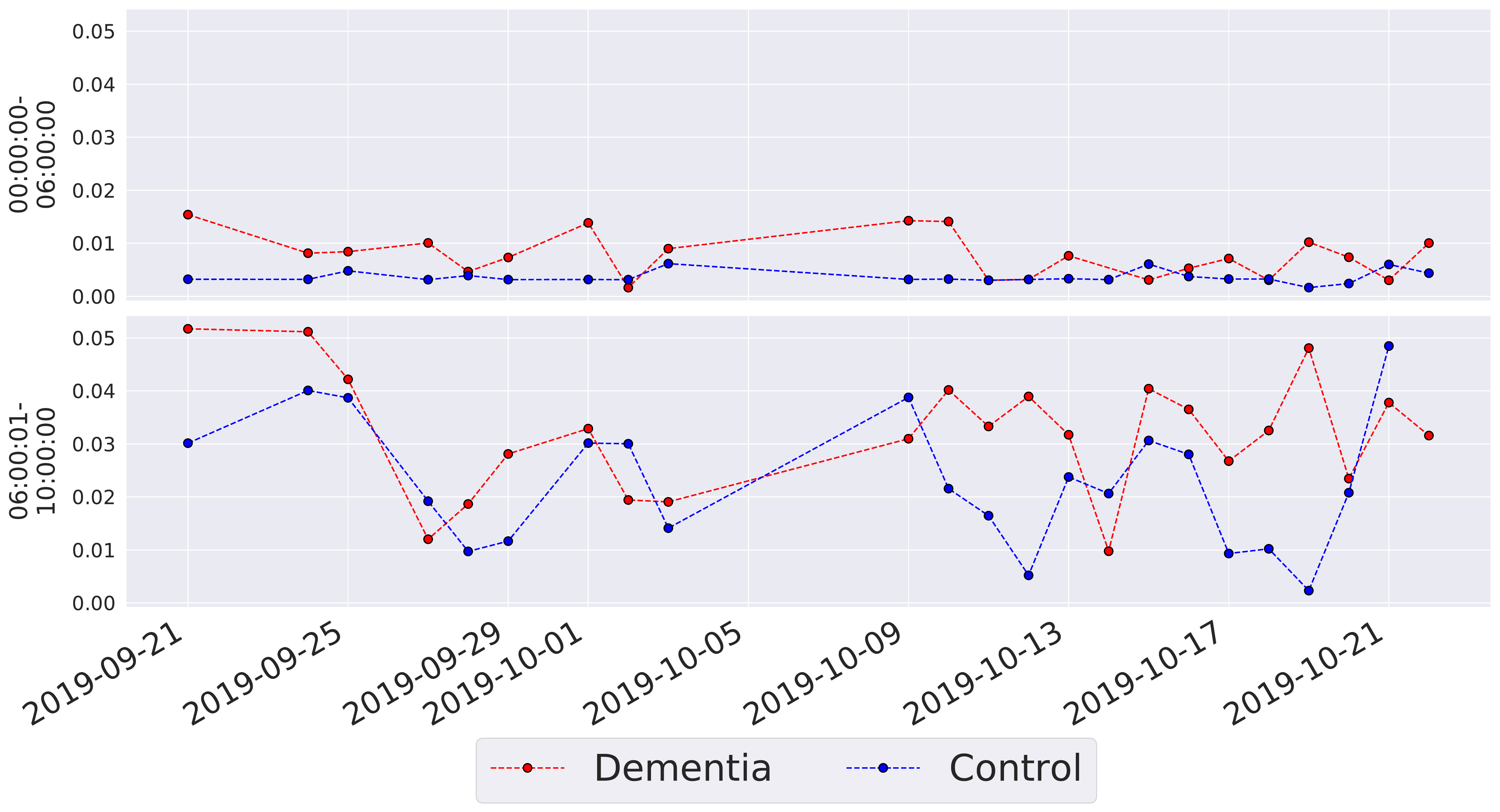}
  \caption{House B}
  \label{fig:entropy5110}
\end{subfigure}
\caption{The complexity of localisation.}
\label{fig:complexities}


    \centering
    \includegraphics[width=0.6\linewidth]{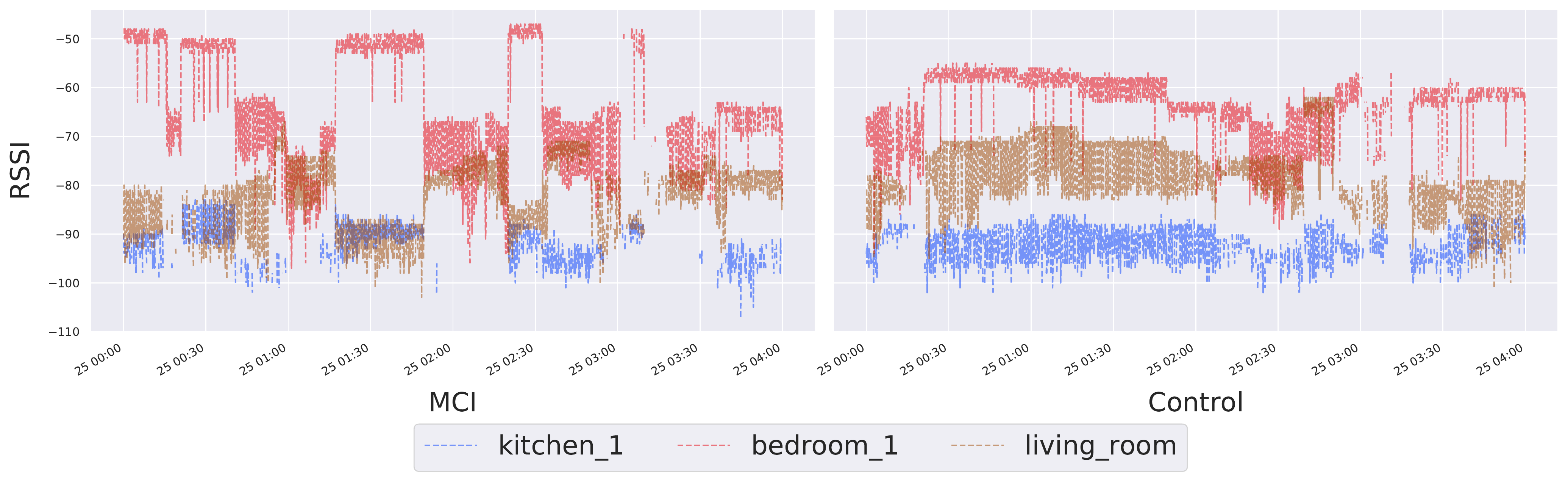}
\caption{Visualisation of sleep disturbance for House A, and PwD (left), partner (right)}
\label{fig:sleeping}
\end{figure*}

The mutual information (MI) between the location of the two residents in each house is measured to estimate shadowing, following a similar approach to that described in \cite{twomey2017unsupervised}. Additionally, in  our analysis we stratify MI by time of the day in order to understand whether significant MI is confined to particular intervals, e.g. morning Fig.\ref{fig:mi}.

\begin{figure*}[t]
    \centering
    \includegraphics[width=0.6\linewidth]{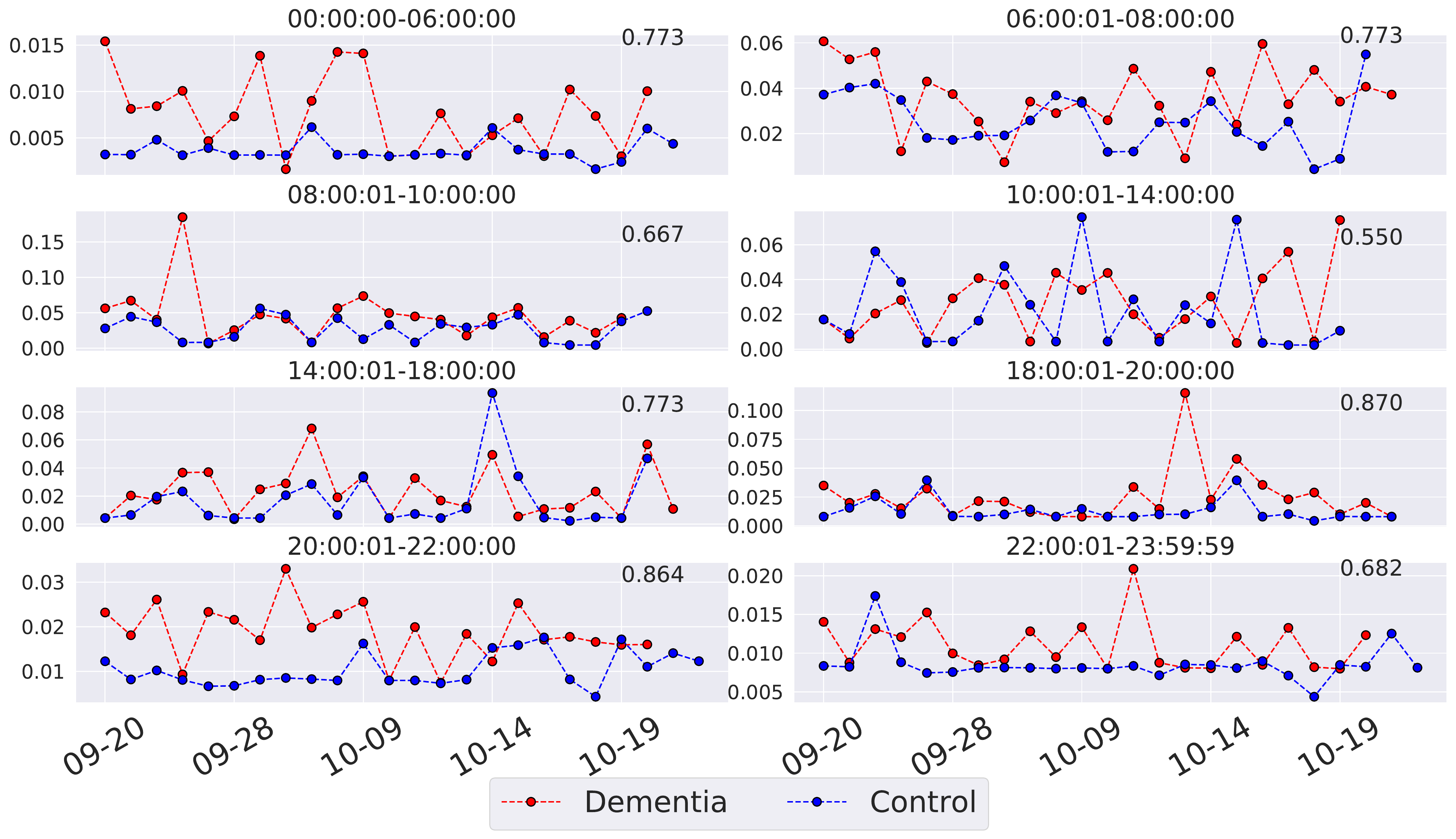}
    \caption{House B - Comparing complexities of localisation.}
    \label{fig:entropies_extra}
\end{figure*}

Temporal lags were introduced in order to allow for short delays in shadowing, with no significant change, and are therefore not included in this work. This is likely due to the fact that the lag interval is significantly shorter than the analysis window. Looking at Fig.\ref{fig:mi}, in the case of house B there is no trend of increasing/decreasing correlation. Even though in the case of house A, there is a slight increase over time, stronger conclusions will be drawn when more data arrives. Our data collection is ongoing, and as it grows we will identify finer, more informative signals of disease state. Time lags are likely to become more relevant with finer and more targeted analysis that may be motivated by changes in behaviour as the disease progresses. 

\subsection{Wandering}


%
The approach for modelling wandering is motivated by its antonym: non-wandering is typified by low activity levels and persistent location predictions. Thus we jointly compare the activity levels and the location complexity of residents in this section. Specifically, in Fig. \ref{fig:activities} we track the total activity per room, as well as the total activity of the day, over the span of the data, and in Fig. \ref{fig:complexities} we track the complexity of the localisation predictions, derived from the \emph{Lempel-Ziv} \cite{lempel1976complexity} complexity measure. It is clear in Fig. \ref{fig:acc5110} that the activity levels of the patient are higher than that of the other participant in this house, and Fig. \ref{fig:entropy5110} gives similar results, although in Fig. \ref{fig:acc3011} and Fig. \ref{fig:entropy3011} it is less obvious, for which there might be several reasons. One explanation could be that the data we have so far for house A spans only a short time period (less than 10 days). A second cause might be that the patient in house A only has MCI, whose daily life only gets affected to a mild degree compared to the participant with dementia in house B.

Additionally, Fig. \ref{fig:entropies_extra} shows in detail the complexity of localisation predictions in different time intervals over the experiment span in house B. On every subplot, the number at the top-right is the average number of times the patient has a higher complexity than the partner. In this analysis, we only include house B, as the span is longer. In the case of house A these averages are not so much in favour of the patient - $5/8$ segment as compared to $8/8$ for house B.

\subsection{Sleep Disturbance}
In both houses, we observe that during the early hours of the day, 0-6 am, the participants have a higher activity levels, while the control/carers have a more steady, and at most times, lower activity levels. This might be an indication of the patients having a more uneasy sleep. As a number of studies suggest \cite{mander2016sleep, hita2012disturbed}, people have MCI or dementia will suffer from different levels of sleep disturbance. And a community-based study \cite{hope1990nature} indicates that nearly one-third of their subjects were reported as having been inappropriately active at night.

As can be seen from Fig. \ref{fig:acc5110}, the line representing the activity levels in bedroom 2 of the participant with dementia is quite unstable compared to that of the other participant during 0-6 am. Although there is not a significant difference between activity levels in bedroom 1 of the two inhabitants during 0-6 am in Fig. \ref{fig:acc3011}, it is because that the person with MCI got up earlier than $6$ am, especially on July 23rd, 24th and 25th, and started walking in the house while the partner was sleeping, which may suggest the increased sleep disturbance in the person with cognitive disorder from another angle.

For reference, we also include a segment of the \emph{RSSI} data in Fig. \ref{fig:sleeping}, from House A, for both participants, from the early hours of the day. 
It highlights the difference in sleeping quality, as observed through RSSI. The step changes are a result of rotations during sleep. It also shows how different sleeping positions can lead to missing values in certain gateways, e.g. \emph{kitchen\_1} at \emph{02:00} and at \emph{03:15}.


\section{Conclusion and Future Work} \label{conclusion}
This paper presents our preliminary results and shows the potential of detecting early symptoms of cognitive disorders automatically by utilizing data acquired from patients' daily activities and signal processing methods equipped with machine learning techniques. We build a machine learning pipeline that considers the difference in training and testing distributions using \emph{MMD}. We account for the sequential nature of data using a \emph{CRF}, and use our domain knowledge to enhance our training data with data coming from the participants. 

One of the limitations of this study is that our localisation precision can only reach room-level for now due to the configuration and the deployment of our sensor system. Finer-grained localisation will expose behavioural reactions to furniture placements across rooms. 

Potential directions for improvements include taking into consideration environmental sensors, such as Passive Infrared and Video, through machine learning techniques such as multi-modal data fusion \cite{diethe2017probabilistic}. Moreover, we will enhance our training data using clinical data through our domain knowledge of activities of daily living.

The paper as it stands establishes baseline activity, sleep and co-localisation dependence that will be used after data collection completed. That we see signals here is promising for our future data collection, analysis and research.

\section*{Acknowledgements}
This research was funded by CUBOID (UK MRC Momentum grant MC/PC/16029) and  the SPHERE IRC (grant EP/K031910/1).

\newpage
\bibliography{ecai}
\balance

\end{document}